\documentclass[aps,twocolumn,floatfix,showpacs,amsmath,amsfonts]{revtex4}

\usepackage{graphicx}
\usepackage{bm}
\usepackage{array}
\usepackage{hyperref}

\begin{document}

\title{Resonance wave functions located at the Stark saddle point}

\author{Holger Cartarius}
\email{Holger.Cartarius@itp1.uni-stuttgart.de}
\author{J\"org Main}
\author{Thorsten Losch}
\author{G\"unter Wunner}
\affiliation{Institut f\"ur Theoretische Physik 1, Universit\"at Stuttgart,
  70550 Stuttgart, Germany}
\date{\today}

\begin{abstract}
We calculate quantum mechanically exact wave functions of resonances in
spectra of the hydrogen atom in crossed external fields and prove the
existence of long-lived decaying quantum states localized at the Stark
saddle point. A spectrum of ground and excited states reproducing the nodal
patterns expected from simple quadratic and cubic expansions of the potential
in the vicinity of the saddle point can be identified. The results demonstrate
the presence of resonances in the vicinity of the saddle predicted by simple
approximations.
\end{abstract}


\maketitle

\section{Introduction}
\label{sec:intro}
The hydrogen atom in crossed static electric and magnetic fields is an 
important example for a quantum system accessible both in experiments and
numerical calculations. Adding external fields to the Coulomb potential of the
hydrogen atom opens the possibility for wave functions to be localized far away
from the nucleus. The existence of such localized states in quantum systems has
attracted the attention of both theoretical \cite{Bha82a,Gay79a,Rau90a,Cla85}
and experimental \cite{Fau87a,Rai93a} investigations over a long period of
time. First investigations considered a gauge dependent variant of
the system, in which the paramagnetic term was omitted 
\cite{Bha82a,Gay79a,Fau87a} due to its smallness as compared with the
diamagnetic term at large field strengths. The resulting sum potential
possesses a minimum away from the nucleus in which localized states can exist.
However, if one takes into account all terms of the magnetic field, one can
immediately see that the outer potential minimum does not exist in the real
physical system, rather, one finds classical electron motion close to the
Stark saddle point which is bound in two directions and unbound in the third
direction. In fact, localized resonance quantum states in the vicinity of the
saddle point have been predicted by Clark et al.\ \cite{Cla85} as quantized
version of this quasi-bound motion.

The question whether or not resonances located at the Stark saddle point
do exist has also a crucial meaning for the ionization mechanism. Classical
electron paths describing an ionization must pass the vicinity of the
Stark saddle. This process was investigated, e.g., using the
transition state theory \cite{Jaf99,Jaf00,Wig01,Uzer02,Waa08}. The transition
state theory is applicable to many dynamical systems which evolve from an
initial to a final state and has its origin in the calculation of chemical
reaction rates. The concept is based on classical trajectories which describe
a reaction by passing a boundary in phase space, viz.\ the transition state,
which must be crossed by all trajectories connecting the initial
(``reactants'') and final (``products'') side. For the hydrogen atom in crossed
fields the theory postulates classical orbits confined in the vicinity of the
Stark saddle point \cite{Uzer02}.

The work by Clark et al.\ \cite{Cla85} can be regarded as a first step
in discussing the ionization mechanism of the hydrogen atom in crossed
electric and magnetic fields in this context. The quasi-bound states confined
to the vicinity of the saddle found by Clark et al.\ coincide with an 
approximation to the Hamiltonian in the framework of the transition state
theory, however, only an algorithmic procedure based on a normal-form
representation of a power-series expansion of the Hamiltonian allowed for
identifying the transition state \cite{Wig01,Uzer02,Waa08}. 
Even though the transition state for the hydrogen atom in crossed fields has
been found, the question of whether or not the classical trajectories in
its vicinity leave a signature in the exact quantum spectrum remained
unanswered for a long time. Finally, clear evidence for signatures in the
energies, i.e., the eigenvalues of the quantum resonances, was found
\cite{Car09a}.

It is the purpose of this paper to demonstrate that there is, indeed, a
relationship between the exact quantum resonances and the quantized energy
levels of the electron motion near the transition state due to a definite
spatial restriction of the resonances' probability density to a small region
around the saddle. We calculate the position space representation of the
corresponding wave functions and are able to show that they are
clearly located in a vicinity of the Stark saddle point. From a quadratic
approximation of the potential around the saddle point one expects a harmonic
oscillator spectrum of energy levels and wave functions. We even find nodal
patterns in the exact quantum states which correspond to ``excited'' states of
the quadratic approximation, and thus demonstrate the usefulness
of the simple approximation and the existence of the resonances predicted by a
classical treatment of the system \cite{Cla85,Uzer02}.

In Sec.\ \ref{sec:system} we introduce the system as well as the second-
and third-order power-series expansions of the potential in the vicinity
of the Stark saddle point. Position space representations of exact quantum
wave functions which prove the existence of resonances located closely
at the position of the saddle are presented in Sec.\ \ref{sec:results}.
Conclusions are drawn in Sec.\ \ref{sec:conclusion}.

\section{Hamiltonian and approximations in the vicinity of the Stark
  saddle point}
\label{sec:system}

\subsection{Hamiltonian and exact quantum calculations}
In Hartree units the Hamiltonian of a hydrogen atom in crossed
external fields with an electric field $f$ orientated along the $x$-axis
and a magnetic field along the $z$-axis represented by the vector potential
$\bm{A}$ has the form
\begin{equation}
  H = \frac{1}{2} \left ( \bm{p} + \bm{A} \right )^2 - \frac{1}{r} + f x .
  \label{eq:Hamiltonian_exact}
\end{equation} 
The parity with respect to reflections at the ($z=0$)-plane is a good quantum
number and allows considering states with even and odd $z$-parity separately.

It is a common and efficient method to rewrite the Schr\"odinger equation for
numerical calculations in dilated semiparabolic coordinates \cite{Mai94,Car09a},
\begin{equation}
  \mu = \frac{1}{b}\sqrt{r+z} ,\quad \nu = \frac{1}{b}\sqrt{r-z} , \quad
  \varphi = \arctan \frac{y}{x} ,
\end{equation}
which yields
\begin{multline}
  \bigg \{ \Delta_\mu + \Delta_\nu - \left ( \mu^2 + \nu^2 \right ) 
  + b^4 \gamma \left ( \mu^2 + \nu^2 \right ) \mathrm{i}\frac{\partial}{\partial
    \varphi} \\ 
  - \frac{1}{4} b^8 \gamma^2 \mu^2\nu^2 \left ( \mu^2 + \nu^2 \right ) 
  - 2 b^6 f \mu\nu \left ( \mu^2 + \nu^2 \right ) \cos \varphi \bigg \} \psi\\ =
  \left \{ -4b^2 + \lambda \left ( \mu^2 + \nu^2 \right ) \right \} \psi 
  \label{eq:Schroedinger_transformed}
\end{multline}
with 
\begin{equation}
  \Delta_\varrho = \frac{1}{\varrho} \frac{\partial}{\partial\varrho} \varrho
  \frac{\partial}{\partial\varrho} + \frac{1}{\varrho^2} 
  \frac{\partial^2}{\partial\varphi^2} , \qquad \varrho \in \left \{ \mu,
    \nu \right \} ,
\end{equation}
and the generalized eigenvalues $\lambda = -(1+2b^4 E)$, which are related to
the energies $E$ of the quantum states. The calculation of the resonances
is done with a diagonalization of a matrix representation of the Schr\"odinger
equation \eqref{eq:Schroedinger_transformed} and the complex rotation method
\cite{Rei82,Del91,Moi98a}. The necessary complex scaling of the coordinates
$\bm{r}$ is introduced via the complex convergence parameter
\begin{equation}
  b = |b| \mathrm{e}^{\mathrm{i}\vartheta/2} .
\end{equation}
Resonances appear as complex eigenvalues $E$, where the real part of $E$
represents the energy and the imaginary part is related to the width
$\Gamma = -2\mathrm{Im}(E)$. 

An adequate complete basis for the matrix representation of the Schr\"odinger
equation \eqref{eq:Schroedinger_transformed} is given by 
\begin{equation}
  | n_\mu, n_\nu, m \rangle = | n_\mu, m \rangle \otimes | n_\nu, m \rangle ,
\end{equation}
where $|n_\varrho m\rangle$ are the eigenstates of the two-dimensional 
harmonic oscillator. The position space representation in dilated
semiparabolic coordinates has the form
\begin{subequations}
  \begin{multline}
    \psi_{n_\mu n_\nu m}(\mu,\nu,\varphi) = 
    \sqrt{\frac{[(n_\mu-|m|)/2]!}{[(n_\mu+|m|)/2]!} 
      \frac{[(n_\nu-|m|)/2]!}{[(n_\nu+|m|)/2]!}} \\
    \times \sqrt{\frac{2}{\pi}} f_{n_\mu m}(\mu) f_{n_\nu m}(\nu)
    \mathrm{e}^{\mathrm{i} m \varphi}
    \label{eq:wave_position}
  \end{multline}
  with the ``radial'' wave functions
  \begin{equation}
    f_{nm}(\varrho)= \mathrm{e}^{-\varrho^2/2} 
    \varrho^{|m|} \mathrm{L}_{(n-|m|)/2}^{|m|} (\varrho^2)
  \end{equation}
\end{subequations}
expressed in terms of Laguerre polynomials $\mathrm{L}_n^\alpha(x)$. Due to
the complex scaling $\mu$ and $\nu$ become complex coordinates. In properly
complex conjugated wave functions obtained with the complex rotation method
one has to bear in mind that only the intrinsically complex parts have to be
conjugated \cite{Res75a,Moi98a}, i.e., in the complex conjugated version
$\psi^\ast_{n_\mu n_\nu m}$ of the wave function \eqref{eq:wave_position}
$\mathrm{e}^{\mathrm{i} m \varphi}$ is replaced with
$\mathrm{e}^{-\mathrm{i} m \varphi}$ and the ``radial'' parts 
$f_{n_\mu m}(\mu)$, $f_{n_\nu m}(\nu)$, which become complex \emph{only} by the
complex scaling, are \emph{not} conjugated. 

It is very effective to perform the matrix setup of the Schr\"odinger equation
\eqref{eq:Schroedinger_transformed} completely algebraically by expressing
Eq.\ \eqref{eq:Schroedinger_transformed} in terms of harmonic oscillator
creation and annihilation operators. Consequently, the expansion coefficients
$c_{i n_\mu n_\nu m}$ of the eigenstates obtained in a matrix diagonalization
\begin{subequations}
  \begin{align}
    \Psi_i(\mu,\nu,\varphi) &= \sum_{n_\mu,n_\nu,m} c_{i n_\mu n_\nu m}
    \psi_{n_\mu n_\nu m} (\mu,\nu,\varphi), \label{eq:eigenstate} \\
    \Psi_i^\ast(\mu,\nu,\varphi) &= \sum_{n_\mu,n_\nu,m} c_{i n_\mu n_\nu m}
    \psi_{n_\mu n_\nu m}^\ast (\mu,\nu,\varphi) \label{eq:eigenstate_conj}
  \end{align}
\end{subequations}
belong to position space wave functions \eqref{eq:wave_position}, which are
normalized eigenstates of two coupled two-dimensional harmonic oscillators,
\begin{multline}
  \int\limits_{0}^{\infty} \mathrm{d}\mu  \int\limits_{0}^{\infty} \mathrm{d}\nu 
  \int\limits_{0}^{2\pi} \mathrm{d} \varphi \:
  \mu\nu \: \psi^*_{n_\mu n_\nu m}(\mu,\nu,\varphi)  
  \psi_{n_\mu^\prime n_\nu^\prime m^\prime}(\mu,\nu,\varphi)    \\
   = \delta_{n_\mu n_\mu^\prime} \delta_{n_\nu n_\nu^\prime} \delta_{m m^\prime} ,
\end{multline}
but are not orthogonal and not normalized in the physical position space
$\bm{r} = (x,y,z)^\mathrm{T}$. The eigenstates \eqref{eq:eigenstate}
and \eqref{eq:eigenstate_conj} of the Schr\"odinger equation 
\eqref{eq:Schroedinger_transformed}, however, can be normalized in such a way
that
\begin{multline}
  \int \mathrm{d}^3 \bm{r} \: \Psi_i^\ast(\mu,\nu,\varphi) 
  \Psi_j(\mu,\nu,\varphi) \\ 
  = b^6  \int\limits_{0}^{\infty} \mathrm{d}\mu  \int\limits_{0}^{\infty} 
  \mathrm{d}\nu \int\limits_{0}^{2\pi} \mathrm{d} \varphi \:
  \mu\nu (\mu^2+\nu^2) \: \Psi_i^\ast \Psi_j = \delta_{ij} .
\end{multline}

Since the proper complex conjugation of the wave functions effects only
intrinsically complex parts, their square moduli must be replaced with
$\Psi_j^\ast \Psi_j$ as introduced in Eqs.\ \eqref{eq:eigenstate} and
\eqref{eq:eigenstate_conj} to obtain a measure for the probability density. For
wave functions of decaying states this product will be complex, and we will
visualize the modulus $|\psi^\ast_{n_\mu n_\nu m} \psi_{n_\mu n_\nu m}|$ in
Sec.\ \ref{sec:results} to show where the resonance wave functions are located.

\subsection{Approximations in the vicinity of the saddle point}

\begin{figure}[tb]
  \centering
  \includegraphics[width=\columnwidth]{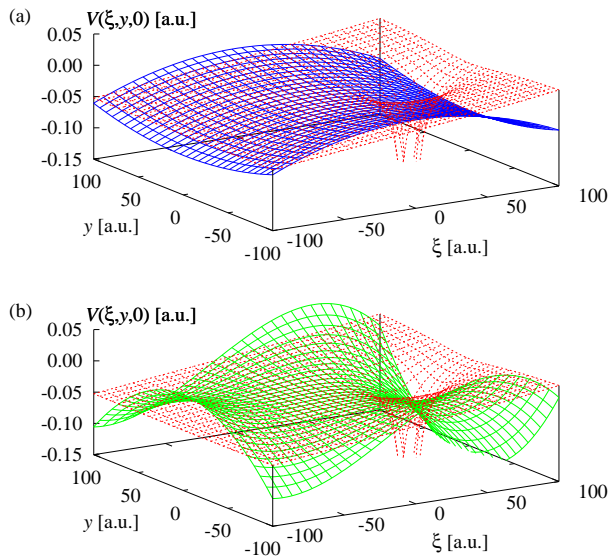}
  \caption{\label{fig:saddle}(Color online) Potential of the
    hydrogen atom in an external electric field (dashed red grids in (a) and
    (b)) in comparison with the second-order (a) and third-order (b)
    power-series expansions around the saddle in the ($\xi$,$y$)-plane with
    $\xi = x-x_s$. Both approximations reproduce the potential the vicinity of
    the saddle at the origin correctly, however, both are only valid very close
    to the saddle point.}
\end{figure}

To investigate resonances located at the saddle point, the electric potential
of the combined contributions of the nucleus and the external electric field, 
\begin{equation}
  V_f = -\frac{1}{r} + f x ,
  \label{eq:pot_exact}
\end{equation}
is expanded into a series up to third-order. As was shown in  Ref.\ 
\cite{Car09a}, already the second-order approximation leads to a good
one-to-one correspondence with energies of exact resonances of the full
Hamiltonian \eqref{eq:Hamiltonian_exact}. Here, we additionally consider an
approximation including all third-order terms to estimate the quality of the
approximation carried out in the vicinity of the saddle point by comparing the
second- and third-order results. The saddle point has the coordinates
$\bm{r}_s  = ( -1/\sqrt{f},0,0)^{\mathrm{T}}$ and its energy has the value
$V_f(\bm{r}_s) = -2\sqrt{f}$. Using the coordinate shift $\xi = x-x_s$ a
power-series expansion around $\bm{r}_s$ yields
\begin{multline}
  V_f(\bm{r}) = -2\sqrt{f} - \sqrt{f}^3 \xi^2 + \frac{1}{2} \sqrt{f}^3
  \left (y^2 + z^2 \right ) \\ - f^2 \xi^3 + \frac{3}{2} f^2 \xi (y^2+z^2)
  + \mathrm{O}((\bm{r}-\bm{r}_s)^4) .
  \label{eq:potential_third order}
\end{multline}
In Fig.\ \ref{fig:saddle}
the second- and third-order power-series expansions around the saddle point
and the full electric potential \eqref{eq:pot_exact} are compared in the
($\xi$,$y$)-plane. The saddle structure is clearly visible and the figure
demonstrates that both approximations are only valid very close to the saddle
point, and thus can describe resonances correctly only in its vicinity.

The approximated Hamiltonian close to $\bm{r}_s$ reads
\begin{multline}
  H = \frac{1}{2} \left ( p_\xi^2 + p_y^2 + p_z^2 \right )
  - \gamma y p_\xi + \frac{1}{2}\gamma^2 y^2 \\
  - 2\sqrt{f} + \frac{1}{2}\sqrt{f}^3\left ( y^2 + z^2 
    - 2 \xi^2\right ) \\ - f^2 \xi^3 + \frac{3}{2} f^2 \xi (y^2+z^2)
  + \mathrm{O}((\bm{r}-\bm{r}_s)^4),
  \label{eq:Hamiltonian_3rd}
\end{multline}
where the gauge $\bm{A} = (-\gamma y,0,0)$ was used as in Refs.\
\cite{Cla85,Uzer02,Car09a}, since it leads to a simple structure of the
terms. Introducing complex scaling parameters $s_i$ via
\begin{subequations}
  \begin{alignat}{2}
    \xi &= Q_\xi/s_\xi , & \quad p_\xi &= s_\xi P_\xi ,\\
    y   &= Q_y/s_y ,    & \quad p_y   &= s_y P_y ,\\
    z   &= Q_z/s_z ,    & \quad p_z   &= s_z P_z 
  \end{alignat}
\end{subequations}
we calculate the resonances of the Hamiltonian with the complex rotation
method. To do so, a matrix representation with a basis of products
\begin{equation}
  | n_\xi, n_y, n_z \rangle =  | n_\xi \rangle \otimes | n_y \rangle \otimes
  | n_z \rangle 
\end{equation}
of one-dimensional harmonic oscillator states $|n_i \rangle$ is built up and
diagonalized.

In the much simpler quadratic approximation, i.e., in the case in which the
two third-order terms $V_3 = -f \xi^3 + 3 f^2 \xi (y^2+z^2)/2$ in Eq.\
\eqref{eq:potential_third order} are neglected, the approximated Hamiltonian
yields a harmonic oscillator spectrum
\begin{multline}
  E_{n_z,n_1,n_2} = -2\sqrt{f} + \omega_z\left (n_z+\frac{1}{2} \right ) \\
  + \omega_1 \left (n_1+\frac{1}{2} \right ) + \omega_2 \left (n_2+\frac{1}{2}
    \right ) ,
\end{multline}
where all frequencies $\omega_i$ can be calculated analytically 
\cite{Cla85,Uzer02,Car09a}. It includes one inverted oscillator, which leads
to a purely imaginary frequency $\omega_2$, and thus represents the resonance
character of the states. The frequency $\omega_z$ belongs to the $z$ motion of
the electron, whereas $\omega_1$ and $\omega_2$ belong to new variables
introduced  via a canonical transformation \cite{Uzer02} to separate the $x$
and $y$ motions.

\section{Results and discussion}
\label{sec:results}

\begin{figure}[tb]
  \centering
  \includegraphics[width=\columnwidth]{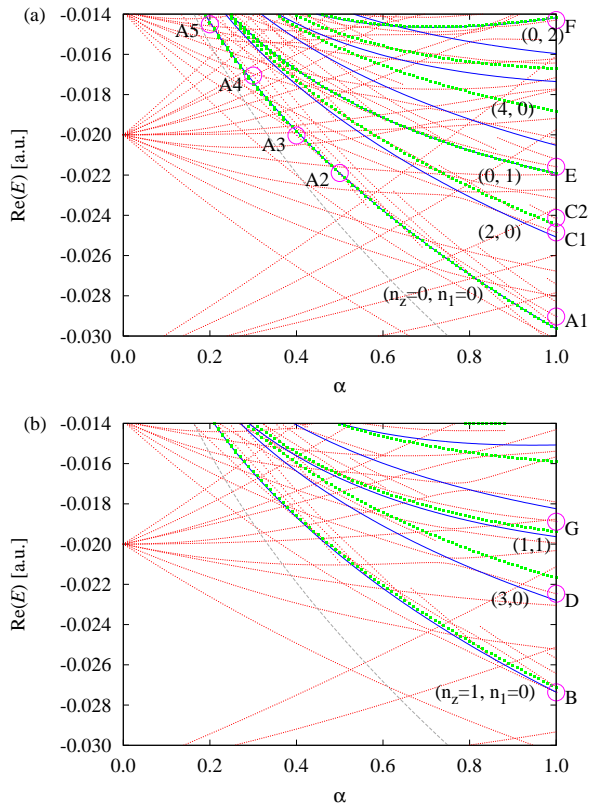}
  \caption{\label{fig:real_energies}(Color online) Comparison of the
  resonance energies (real parts of the complex energy eigenvalues) 
  obtained in exact solutions of the full Hamiltonian 
  \eqref{eq:Hamiltonian_exact} (red dotted lines) and the second- (solid blue
  lines) and third-order (filled green squares) approximations. The results
  are shown for even (a) and odd (b) $z$ parity separately. Position space
  representations of the exact quantum resonances labeled by capital letters
  are shown in Figs.\ \ref{fig:resonance_000_10} to 
  \ref{fig:resonance_n0m_10}. In (a) and (b) the dashed grey lines represent
  the saddle point energy.}
\end{figure}

In Ref.\ \cite{Car09a} it was shown that the energies of some of the
resonances calculated in the quadratic approximation around the saddle point
agree very well with the exact quantum energies over a large range in the
parameter space. The region considered was defined by lines, i.e.,
one-dimensional objects, in the two-dimensional parameter space to obtain
clearly identifiable results. In this paper we chose one of these lines, viz.\
\begin{subequations}
  \begin{align}
    \gamma &= 0.008\times \alpha  , \label{eq:line_1a} \\
    f &= 0.0003 \times \alpha ,  \\
     0 &< \alpha < 1 , \label{eq:line_1c}
  \end{align}
\end{subequations}
on which all parameters used in what follows are located. Figure
\ref{fig:real_energies}
shows how the real parts of the resonance energies behave on this line.
The red dotted lines represent the exact quantum solution of the full
Hamiltonian \eqref{eq:Hamiltonian_exact}, the solid blue lines and
the filled green squares denote the second- and third-order energies,
respectively, and the saddle point energy is marked by the dashed grey
lines. Results are shown for even [Fig.\ \ref{fig:real_energies}(a)]
and odd [Fig.\ \ref{fig:real_energies}(b)] $z$ parity.

As was already pointed out in Ref.\ \cite{Car09a} some of the second-order
resonances are traced by the exact solutions of
\eqref{eq:Hamiltonian_exact}. Here we see that this is in particular true for
all  resonances of the second-order power-series expansion (solid blue lines in
Fig.\ \ref{fig:real_energies}), which have a clear correspondence in the
third-order results (filled green squares in Fig.\ \ref{fig:real_energies}), 
i.e., which have almost the same energies in both approximations for all parameters
$\alpha$. In other words, approximated resonances in the vicinity of the Stark
saddle point whose energies seem already to be converged in the two lowest-order power-series expansions have also a counterpart in the exact spectrum of
the full quantum system described by the Hamiltonian
\eqref{eq:Hamiltonian_exact}. In these cases the simple power-series expansions
seem to provide a good approximation for resonances located at the saddle. Due
to the very good agreement of the second-order and the exact quantum results in
a large region of the parameter space it was already concluded in Ref.\
\cite{Car09a} that the corresponding exact crossed-fields hydrogen atom
resonances must be closely related with localized states predicted by 
Clark et al. 

\begin{figure}[tb]
  \centering
  \includegraphics[width=\columnwidth]{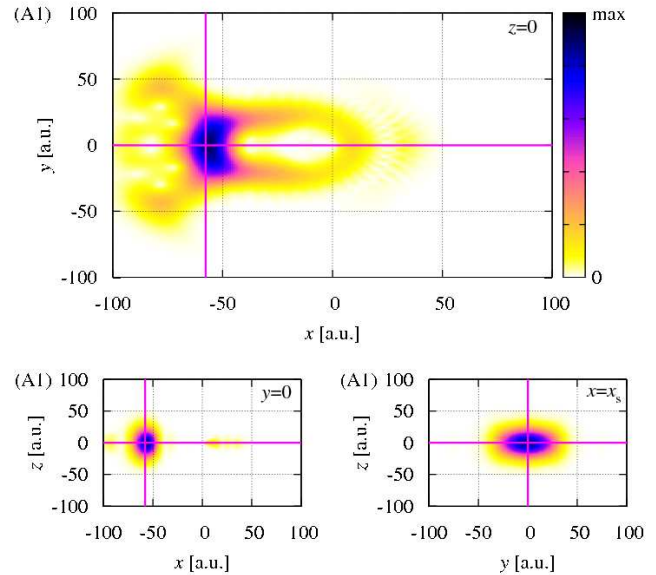}
  \caption{\label{fig:resonance_000_10}(Color online) Density plots of the
    resonance labeled A1 in Fig.\ \ref{fig:real_energies} in the planes $z=0$,
    $y=0$, and $x=x_s$. The solid magenta lines mark the saddle point. One can
    clearly see that the probability density of the resonance is restricted to
    a close vicinity around the saddle point.}
\end{figure}

\begin{figure}[tb]
  \centering
  \includegraphics[width=\columnwidth]{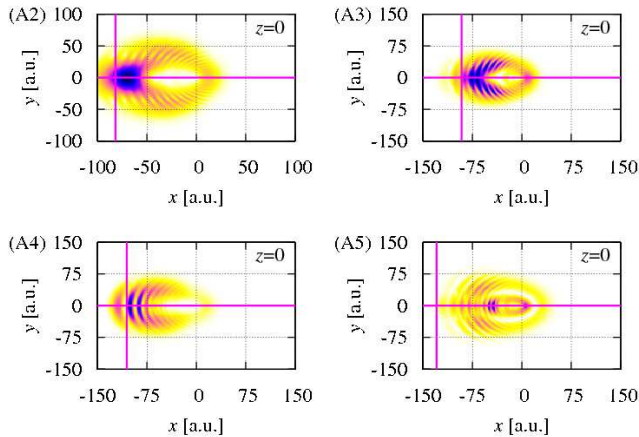}
  \caption{\label{fig:resonance_000_alphas}(Color online) Density plots
    of the resonances labeled A2, A3, A4, and A5 in Fig.\ 
    \ref{fig:real_energies} in the $z=0$ plane. For decreasing field strengths
    one can observe that the resonances are decreasingly centered at the
    saddle point.}
\end{figure}

In this paper we show using the position space representation of the exact
quantum wave functions that they are clearly centered at the Stark saddle
point in all cases, in which the good agreement in the energies described above
appears. The best correspondence is expected for the lowest transition state
resonance with quantum numbers $n_z=0$, $n_1=0$ in the second-order
approximation and for the largest field strengths. The position space wave
function of this resonance for $\alpha = 1$, i.e., $\gamma = 0.008$,
$f = 0.0003$ (labeled A1 in Fig.\ \ref{fig:real_energies}) is shown in Fig.\
\ref{fig:resonance_000_10}
in the planes $z=0$, $y=0$, and $x=x_s$. All three sections show that the
probability density of the resonance is centered at and restricted to a
close vicinity of the saddle point, which is marked by the solid magenta lines.
The transformation of the quantum resonances corresponding to the same
second-order line for decreasing field strengths can be observed in Fig.\ 
\ref{fig:resonance_000_alphas},
where examples for $\alpha = 0.5$, $\alpha = 0.4$, $\alpha = 0.3$, and
$\alpha = 0.2$ are drawn in the $z=0$ plane. Between $\alpha = 1$ 
(resonance labeled A1 in Fig.\ \ref{fig:real_energies}) and $\alpha = 0.5$ 
(A2) the shape of the resonance does not change much, it stays
centered at the saddle point, however, a radially oscillating probability
density at the borders becomes more and more pronounced indicating an
increasing contribution of typical Rydberg states. For $\alpha = 0.4$ (A3) one
can already observe that the whole wave function is affected by the
oscillations.
The influence of the Coulomb potential becomes large enough to lead to a
wave function with nodal lines in radial direction for $\alpha = 0.3$ (A4),
whose main maximum is still located at the Stark saddle point. This
localization is lost if one decreases the field strengths further, which
can be seen for $\alpha = 0.2$ (A5), where the maximum of the probability
amplitude is far away from the saddle.

\begin{figure}[tb]
  \centering
  \includegraphics[width=\columnwidth]{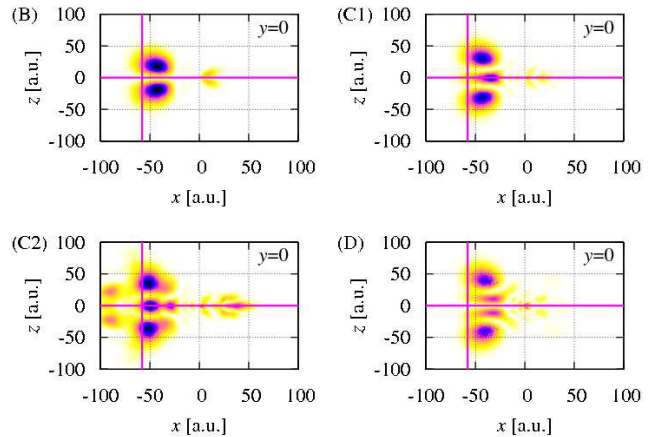}
  \caption{\label{fig:resonance_00n_10}(Color online) Density plots
    of the resonances B, C1, C2, and D (cf.\ Fig.\ \ref{fig:real_energies})
    in the $z=0$ plane. The resonances can be assigned to second-order
    approximations with $n_1 = 0$ and $n_z \neq 0$. The nodal patterns
    expected for excited states in $z$ direction can be observed.}
\end{figure}

There are a number of ``excited'' second-order states which have a
correspondence in the third-order power-series expansion as well as in the
exact quantum spectrum. Figure \ref{fig:resonance_00n_10}
compares resonances with excitations in the $n_z$ quantum number. In the 
second-order approximation these energies correspond to excited harmonic
oscillator states in $z$ direction. The resonance labeled B represents the
quantum number $n_z=1$, i.e., the first excited state. One nodal line
at $z=0$ is expected in the section and can be observed in Fig.\
\ref{fig:resonance_00n_10}. One may argue that a nodal plane at $z=0$ is not
surprising for a state with odd $z$ parity, however, further excitations whose
nodal patterns reproduce the expectations correctly can be found. Comparing the
second-order power-series result for quantum numbers $n_z=2$, $n_1=0$ and its
third-order counterpart in Fig.\ \ref{fig:real_energies}(a) we find two
possibilities to assign exact quantum resonances. Indeed, we find the two
states labeled C1 and C2 whose probability densities are located very close to
the saddle point. Both of these resonances show the two nodal lines expected
for $n_z=2$ and demonstrate that a rich spectrum of states dominated by the
local shape of the potential around the Stark saddle point is present. The
resonances C1 and C2 have almost the same real part of the energy but differ
significantly in their imaginary parts. One can assume that these resonances
can be assigned to two second-order states with $n_z=2$, $n_1=0$ and different
$n_2$ values, which do not influence the real energy in the second-order
approximation. A definite answer, however, is not possible because the excellent
agreement in the real parts of the energies of the approximated states with the
exact quantum results of the full Hamiltonian \eqref{eq:Hamiltonian_exact} does
not hold for the imaginary parts \cite{Car09a}. The imaginary parts
of the energies obtained with the second- and third-order power-series
expansions of the potential do not agree as well as their real parts and cannot
be considered converged in these simple approximations. Their quality does
not allow for a comparison with their numerically exact counterparts. Note
that the exact computations with the complex rotation method provide accurate
results for both the real and imaginary parts of the energies. Finally
one can even observe the resonance labeled D with three nodal lines as one
would expect for a $n_z = 3$ state, and indeed, the resonance energy can be
mapped to that of the second-order state $n_z=3$, $n_1=0$. The line belonging
to resonance $n_z=4$, $n_1=0$ in Fig.\ \ref{fig:real_energies}(a) does not
seem to have a correspondence in the third-order approximation and it was not
possible to find exact quantum resonances with the proper nodal structure.

\begin{figure}[tb]
  \centering
  \includegraphics[width=\columnwidth]{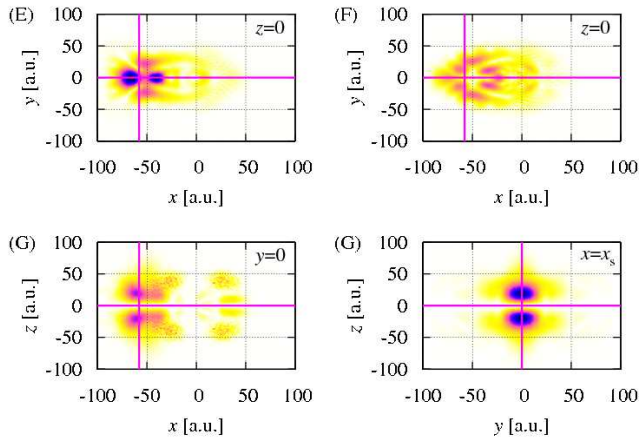}
  \caption{\label{fig:resonance_n0m_10}(Color online) Density plots
    of the resonances E, F, and G (cf.\ Fig.\ \ref{fig:real_energies})
    corresponding to states of the quadratic approximation excited in $n_1$. 
    Complicated nodal patterns are found.}
\end{figure}

Further examples including excitations $n_1 \neq 0$ can be found in Fig.\
\ref{fig:resonance_n0m_10}.
Due to the energy diagram in Fig.\ \ref{fig:real_energies}(a) resonance E can
be assigned to quantum numbers $n_z=0$, $n_1=1$. The position space
representation in Fig.\ \ref{fig:resonance_n0m_10} reveals a complicated
shape of the probability density, which is not very surprising as the
excitation in $n_1$ belongs to a coordinate whose origin is in a canonical
transformation including the positions $x$, $y$ as well as the momenta
$p_x$ and $p_y$. However, the resonance is clearly located close to the
position of the saddle. This is also true for resonance F, which is assigned to
the second-order quantum numbers $n_z=0$, $n_1=2$ and shows even a more
complicated nodal pattern in position space. Resonance G is connected 
with an excitation in both $n_z$ and $n_1$ ($n_z=1$, $n_1=1$) and exhibits
a nodal plane at $z=0$ due to the respective quantum number $n_z = 1$.

\section{Conclusion}
\label{sec:conclusion}

By determining the position space representation of quantum mechanically exact
wave functions we have proved the existence of resonances located in a close
vicinity of the Stark saddle point which were predicted by simple classical
approximations as, e.g., the transition state theory. The results show that
near the saddle one can find a spectrum of several states belonging to the
local neighborhood of the saddle and ignoring the global structure of the 
Coulomb potential. It reproduces the expectations of a simple quadratic
power-series expansion around the saddle in the energies as well as in the shape
of the wave functions. In particular, one can, in the quadratic approximation,
separate three one dimensional harmonic oscillators one of which is connected
with the $z$ direction. In this spatial direction we found nodal planes of
resonances which have the structure of ground and excited states from $n_z = 0$
up to $n_z = 3$. Complicated nodal patterns are found for excitations in the
quantum number $n_1$ not directly connected to a spatial coordinate.

The comparison of the two approximations of the resonance energy eigenvalues
(cf.\ Fig.\ \ref{fig:real_energies}) showed that some of the second-order 
energies have a distinct correspondence in the third-order approximation
while some have not. In all cases in which an assignment of the results of both
approximations is possible we were also able to detect an exact quantum
resonance with a strong localization at the saddle, whereas such a strong
connection to the saddle could not be observed for exact resonances not
belonging to a pair of second- and third-order resonances. Despite this clear
result it must be noted that, as was already pointed out in a previous
publication \cite{Car09a}, the simple power-series expansions are not capable of
reproducing the resonance widths (or imaginary parts of the energy
eigenvalues) correctly. We have demonstrated, however, that a convergence of
the approximated resonance energies (or real parts) is already a strong 
signature of the existence of exact quantum states centered at the saddle.

The results prove that resonances located at an outer saddle predicted in a 
variety of theoretical work \cite{Bha82a,Gay79a,Fau87a,Cla85,Uzer02} and
supported by experimental results \cite{Rai93a} do exist in the Coulomb
potential superimposed by two external fields. The correspondence already
appears for very simple second- and third-order power-series expansions of the
potential at the saddle point. For future work comparisons with more thorough
approximations applicable to the problem, e.g., the normal form expansion
developed for identifying the classical transition state in high-dimensional
systems \cite{Wig01,Uzer02,Waa08} or its quantum analogue \cite{Schu06a,Waa08}
will be of high value.

%\bibliography{paper.bib}

\end{document}